# Researcher

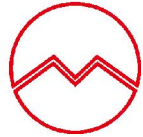

MARSLAND PRESS
Multidisciplinary Academic Journal Publisher

**Design and Control of Steam Flow in Cement Production Process using Neural Network Based Controllers**

Mustefa Jibril[1], Messay Tadese[2], Eliyas Alemayehu Tadese[3]

[1] Msc, School of Electrical & Computer Engineering, Dire Dawa Institute of Technology, Dire Dawa, Ethiopia
[2] Msc, School of Electrical & Computer Engineering, Dire Dawa Institute of Technology, Dire Dawa, Ethiopia
[3] Msc, Faculty of Electrical & Computer Engineering, Jimma Institute of Technology, Jimma, Ethiopia

**Abstract:** In this paper a NARMA L2, model reference and neural network predictive controller is utilized in order to control the output flow rate of the steam in furnace by controlling the steam flow valve. The steam flow control system is basically a feedback control system which is mostly used in cement production industries. The design of the system with the proposed controllers is done with Matlab/Simulink toolbox. The system is designed for the actual steam flow output to track the desired steam that is given to the system as input for two desired steam input signals (step and sine wave). In order to analyze the performance of the system, comparison of the proposed controllers is done by simulating the system for the two reference signals for the system with and without sensor noise disturbance. Finally the comparison results prove the effectiveness of the presented process control system with model reference controller.

**Keywords:** NARMA L2, Model reference controller, Predictive controller

## 1. Introduction

Process engineers are regularly chargeable for the operation of chemical approaches. As these methods turn out to be large scale and/or extra complicated, the position of control automation becomes increasingly essential. To automate the operation of a process, it's far important to use measurements of process outputs or disturbance inputs to make selections approximately the proper values of manipulated inputs.

A chemical- process working unit frequently includes numerous unit operations. The control of a working unit is commonly reduced to considering the manipulated of every unit operation one at a time. Even so, every unit operation may also have a couple of, on occasion conflicting targets, so the development of manipulate goals isn't a trivial trouble.

The closed loop drift control system is basically a remarks control system. Process loop control which utilized in chemical and petrochemical vegetation, oil refineries, metallic plant, cement kilns, paper milling and pharmaceuticals, waste water treatment plant and so forth. The early production system became herbal scale up version of the conventional manual practices. In drift control loop numerous element are used which carry out accurately according their function.

Once the control structure is determined, it's far vital to determine on the manipulate set of rules. The control algorithm uses measured output variable values (alongside desired output values) to exchange the manipulated input variable. A manipulate algorithm has some of control parameters, which have to be "tuned" (adjusted) to have perfect performance. Often the tuning is accomplished on a simulation model earlier than implementing the control strategy on the actual method.

## 2. Mathematical model

The process control system of a cement factory controller which controls the outlet steam a long with terminal variable is shown in Figure 1 below. The input is voltage V (s) and the output is outlet steam Q (s).

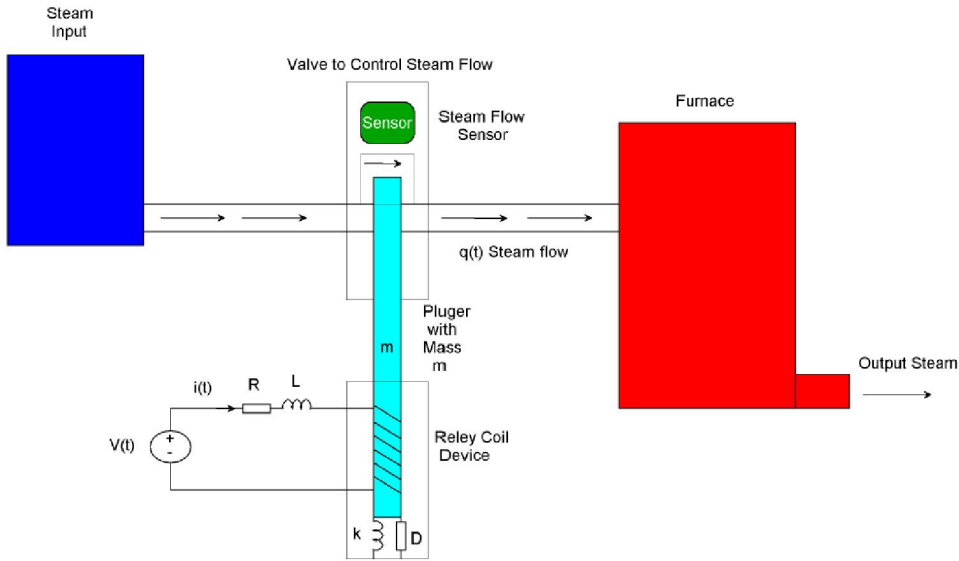

Figure 1 Cement factory process control diagram

For the electrical circuit, V (t) is

$$V(t) = Ri(t) + L\frac{di(t)}{dt} \qquad (1)$$

Taking the Laplace transform of equation (1) yields:

$$V(s) = RI(s) + LsI(s) \qquad (2)$$

The transfer function of input voltage to output current become

$$\frac{I(s)}{V(s)} = \frac{1}{Ls + R} \qquad (3)$$

The mechanical force developed in the relay coil device for the plunger is

$$F(t) = k_m i(t) \qquad (4)$$

Where
Km= Relay constant N/A
The equation of motion of the plunger is

$$F(t) = k_m i(t) = m\frac{d^2x}{dt^2} + D\frac{dx}{dt} + kx \qquad (5)$$

Taking the Laplace transform

$$k_m I(s) = (s^2 m + sD + k)X(s) \qquad (6)$$

The transfer function of input current to output displacement become

$$\frac{X(s)}{I(s)} = \frac{k_m}{s^2 m + sD + k} \qquad (7)$$

For the steam flow

$$q(t) = r(t)x(t) \qquad (8)$$

Where

$$r(t) = pe^{-pt}$$

r (t)= steam flow sensor transfer function
p= sensor sensitivity
The transfer function of the input displacement to the output steam become

$$\frac{Q(s)}{X(s)} = \frac{p}{s + p} \qquad (9)$$

The overall transfer function of the input voltage to the output steam is computed by multiplying equation 3, 7 and 9 yields to

$$\frac{Q(s)}{V(s)} = \frac{pk_m}{(Ls + R)(s + p)(s^2 m + sD + k)} \qquad (10)$$

The parameters of the system is shown in Table 1 below

Table 1 Parameter of the system

| No | Parameters | Symbol | Values |
|----|------------|--------|--------|
| 1 | Inductance | L | 1 H |
| 2 | Resistance | R | 5 ohm |
| 3 | Mass | M | 1 kg |
| 4 | Damper | D | 1N.s/m |
| 5 | Spring | k | 2 N/m |
| 6 | Relay constant | Km | 0.25 N/A |
| 7 | Steam flow sensor sensitivity | p | 3 |

The numerical value of the transfer function will be

$$\frac{Q(s)}{V(s)} = \frac{0.75}{s^4 + 9s^3 + 25s^2 + 31s + 30}$$

## 3.     Proposed Controllers Design
### 3.1     NARMA-L2 Controller Design

One of the primary capabilities of the NARMA-L2 neuro-controller is to transform nonlinear system dynamics into linear dynamics by canceling the nonlinearities. We starts off evolved by means of describing how the identified neural community model may be used to design a controller. The advantage of the NARMA-L2 form is that you may remedy for the control input that reasons the system output to observe a reference signal:

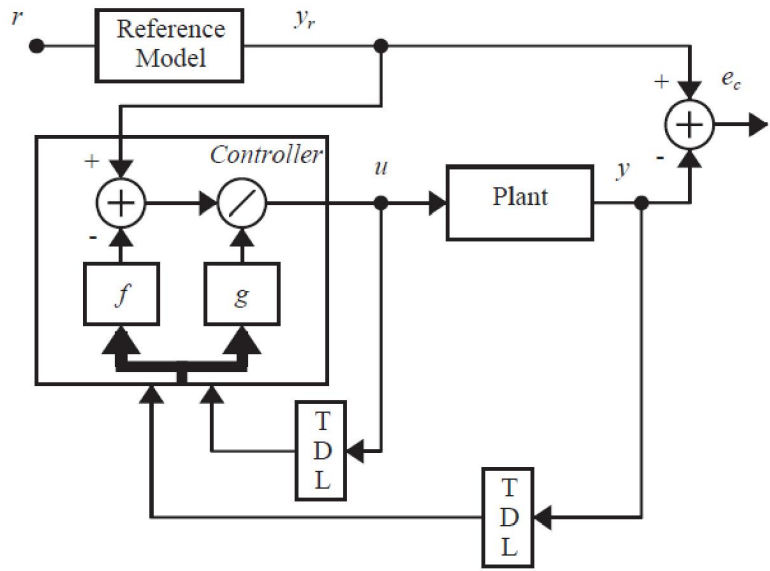

Figure 2 NARMA-L2 Controller.

### 3.2     Model Reference Controller Design

The model reference controller is designed to include two neural networks: a neural network controller and a neural network plant model, as shown in Figure 3. The plant model is diagnosed first, after which the controller is trained in order that the plant output follows the reference model output

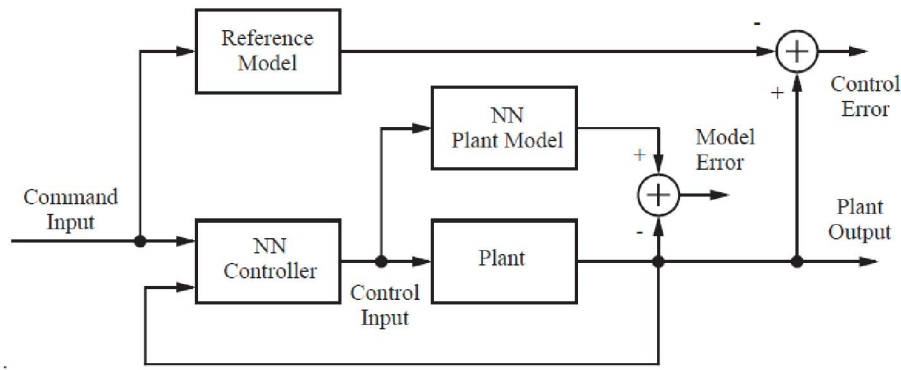

Figure 3 Model Reference Control Architecture

### 3.3     Predictive Controller Design

There are distinctive varieties of neural network predictive controller which can be based on linear model controllers. The proposed neural network predictive controller uses a neural network model of a nonlinear plant to predict destiny plant overall performance. The proposed controller then calculates the manipulated input to be able to optimize plant overall performance over a specific destiny time horizon. The primary goal of the model predictive control is to decide the neural network plant model. Then, the plant model is utilized by the controller to predict destiny overall performance. The technique is represented by using Figure 4.

Table 2 illustrates the network architecture, training data and training parameters of the proposed controllers.

Table 2 Neural network Parameters

| Network Architecture | | | |
|---|---|---|---|
| Size of hidden layer | 6 | Delayed plant input | 4 |
| Sample interval (sec) | 0.1 | Delayed plant output | 4 |
| Training Data | | | |
| Training sample | 65 | Maximum Plant output | 2 |
| Maximum Plant input | 2 | Minimum Plant output | 1 |
| Minimum Plant input | 1 | Max interval value (sec) | 30 |
| Min interval value (sec) | | | 15 |
| Training Parameters | | | |
| Training Epochs | | | 65 |

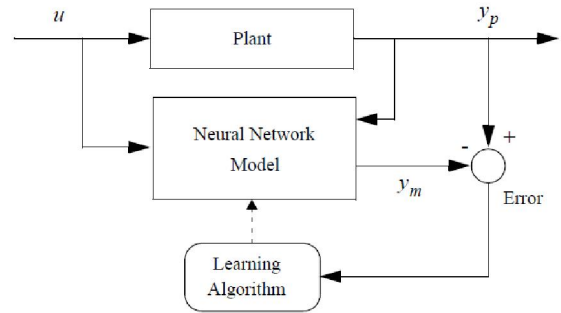

Figure 4 Plant Identification

## 4. Result and Discussion

In this section, comparison of the proposed controllers for tracking the desired steam input signals (step and sine wave) with and without steam flow sensor disturbance will be simulated and analyzed.

### 4.1 Comparison of the Proposed Controllers for Tracking Desired Steam Input Step Signal

The Simulink model of the process control system with the proposed controllers for tracking the desired steam input step signal is shown in Figure 5 below.

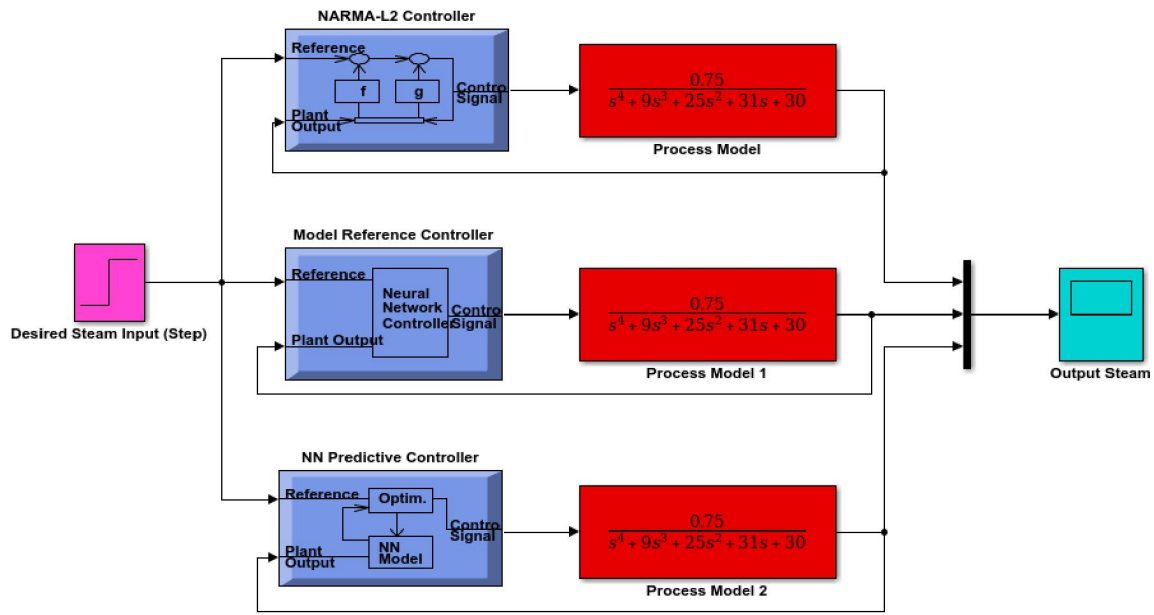

Figure 5 Simulink model of the process control system with the proposed controllers for tracking the desired steam input step signal

The simulation result of the process control system with the proposed controllers for tracking the desired steam input step signal is shown in Figure 6 below.

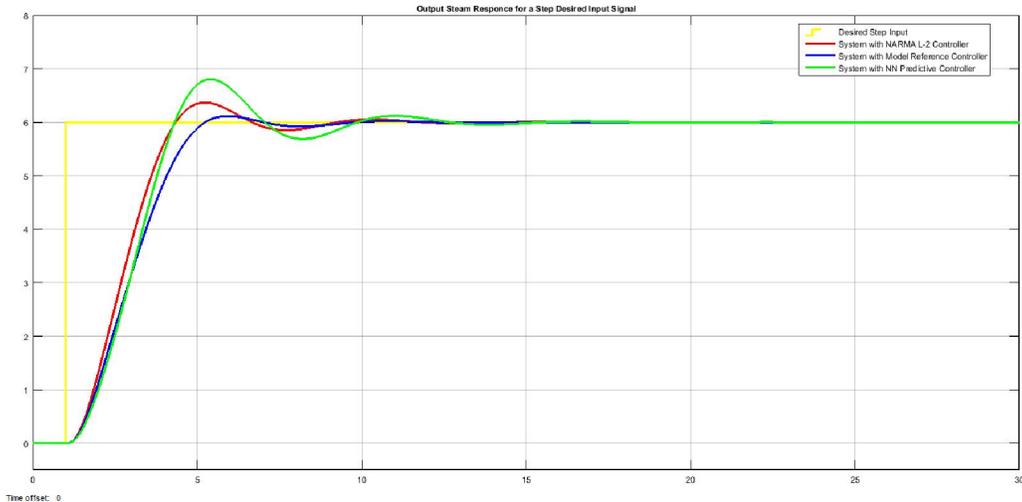

Figure 6 simulation result of the process control system with the proposed controllers for tracking the desired steam input step signal

Table 2 shows the performance characteristic of the simulation result

Table 2 Performance characteristic

| No | characteristic | NARMA L-2 | Model Reference | NN Predictive |
|----|----------------|-----------|-----------------|---------------|
| 1 | Rise time (sec) | 2.4 | 2.45 | 2.45 |
| 2 | Percentage Overshoot (%) | 6 | 1.02 | 13.33 |
| 3 | Settling time (sec) | 11 | 9 | 14.3 |
| 4 | Steady state value | 1 | 1 | 1 |

Table 2 shows that the three controllers have almost the same rise time but the process control system with model reference controller has a small settling time and percentage overshoot as compared to the two proposed controllers.

**4.2　Comparison of the Proposed Controllers for Tracking Desired Steam Input Step Signal with the Presence of Steam Flow Sensor Disturbance**

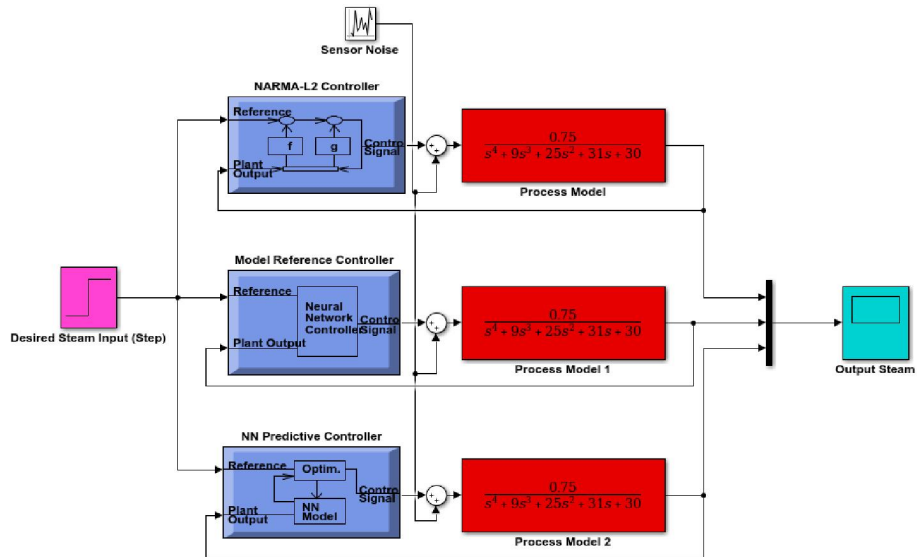

Figure 7 Simulink model of the process control system with the proposed controllers for tracking the desired steam input step signal with the presence of steam flow sensor disturbance

The Simulink model of the process control system with the proposed controllers for tracking the desired steam input step signal with the presence of steam flow sensor disturbance and the sensor Disturbance is shown in Figure 7 and Figure 8 respectively.

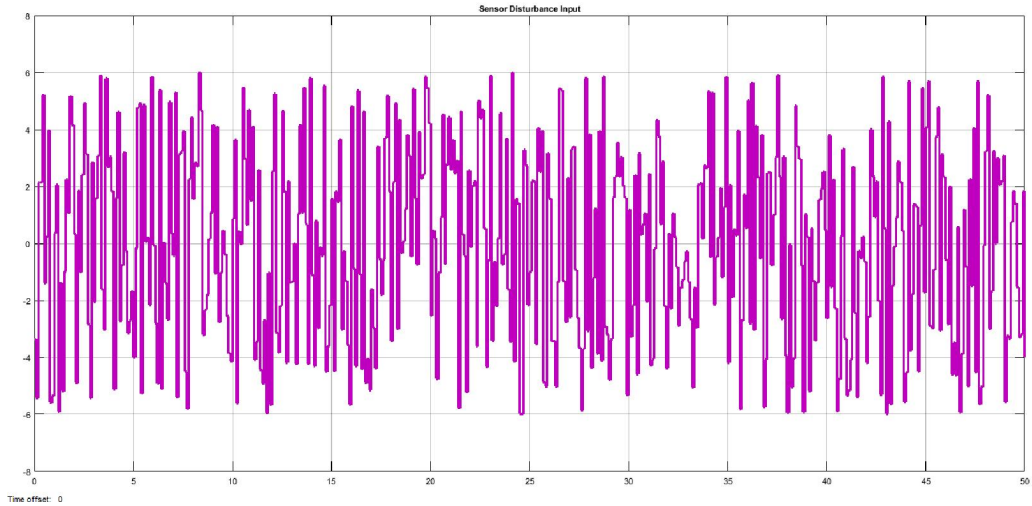

Figure 8 Sensor disturbance

The simulation result of the process control system with the proposed controllers for tracking the desired steam input step signal with the presence of steam flow sensor disturbance is shown in Figure 9 below.

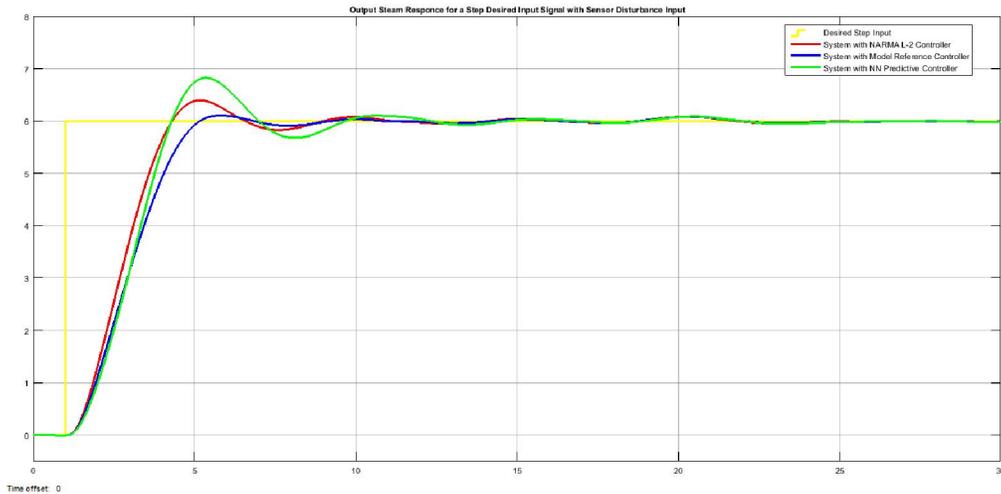

Figure 9 simulation result of the process control system with the proposed controllers for tracking the desired steam input step signal with the presence of steam flow sensor disturbance

Table 3 shows the performance characteristic of the simulation result

Table 3 Performance characteristic

| No | characteristic | NARMA L-2 | Model Reference | NN Predictive |
|---|---|---|---|---|
| 1 | Rise time (sec) | 2.6 | 2.75 | 2.75 |
| 2 | Percentage Overshoot (%) | 8.33 | 3.33 | 15 |
| 3 | Settling time (sec) | 19 | 18 | 25 |
| 4 | Steady state value | 1 | 1 | 1 |

Table 3 shows that the performance characteristic of the three controllers have been changed. But still the process control system with model reference controller has a small settling time and percentage overshoot as compared to the two proposed controllers.

### 4.3    Comparison of the Proposed Controllers for Tracking Desired Steam Input Sine Wave Signal

The Simulink model of the process control system with the proposed controllers for tracking the desired steam input sine wave signal is shown in Figure 10 below.

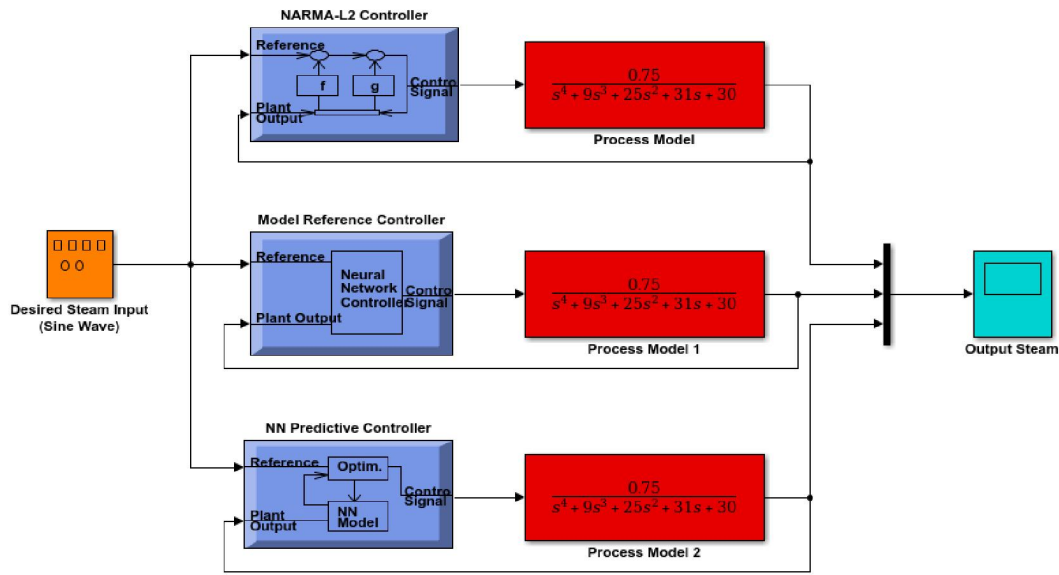

Figure 10 Simulink model of the process control system with the proposed controllers for tracking the desired steam input sine wave signal

The simulation result of the process control system with the proposed controllers for tracking the desired steam input sine wave signal is shown in Figure 11 below.

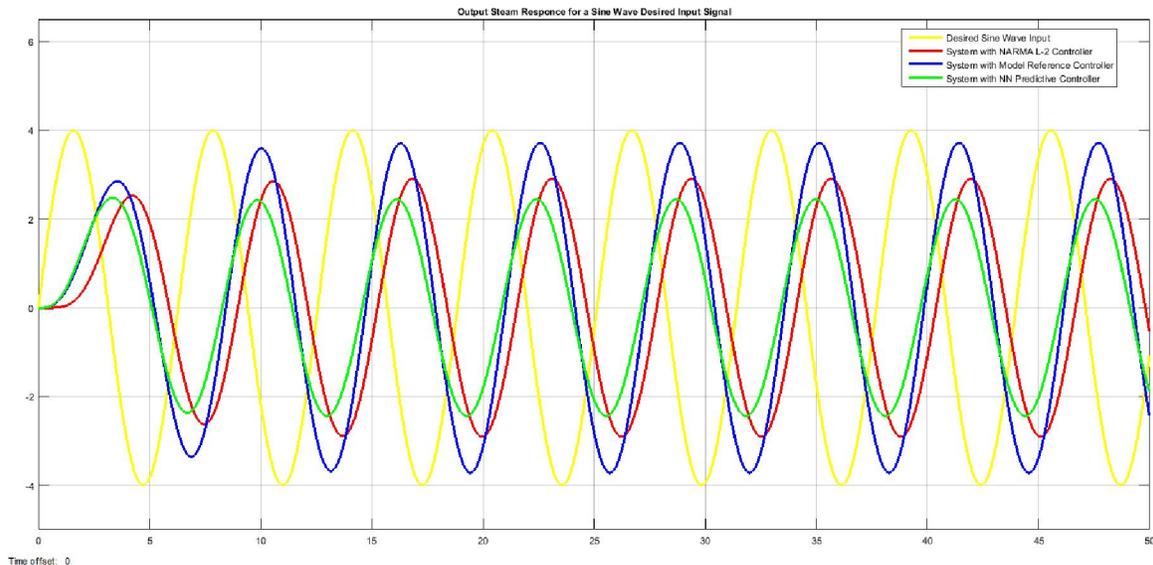

Figure 11 simulation result of the process control system with the proposed controllers for tracking the desired steam input sine wave signal

Table 4 shows the performance characteristic of the simulation result

<div align="center">Table 4 Performance characteristic</div>

| No | characteristic | Peak value (m) |
|---|---|---|
| 1 | Sine wave signal | 4 |
| 2 | NARMA L-2 | 3 |
| 3 | Model Reference | 3.8 |
| 4 | NN Predictive | 2.6 |

Table 4 shows that the process control system with model reference controller have track the desired sine wave signal with 3.8 m peak value as compared to the two proposed controllers.

**4.4 Comparison of the Proposed Controllers for Tracking Desired Steam Input Sine Wave Signal with the Presence of Steam Flow Sensor Disturbance**

The Simulink model of the process control system with the proposed controllers for tracking the desired steam input sine wave signal with the presence of steam flow sensor disturbance is shown in Figure 12 below.

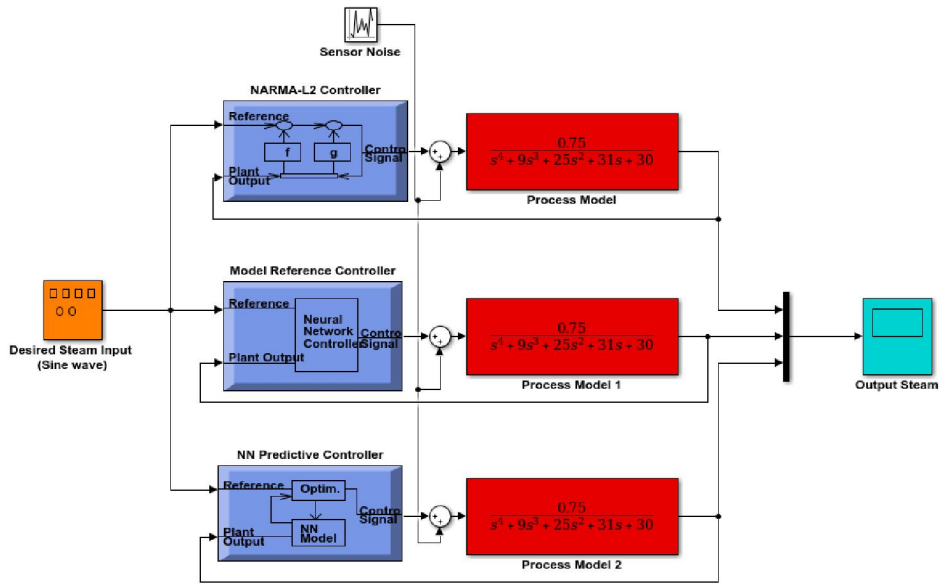

Figure 12 Simulink model of the process control system with the proposed controllers for tracking the desired steam input sine wave signal with the presence of steam flow sensor disturbance

The simulation result of the process control system with the proposed controllers for tracking the desired steam input sine wave signal with the presence of steam flow sensor disturbance is shown in Figure 13 below.

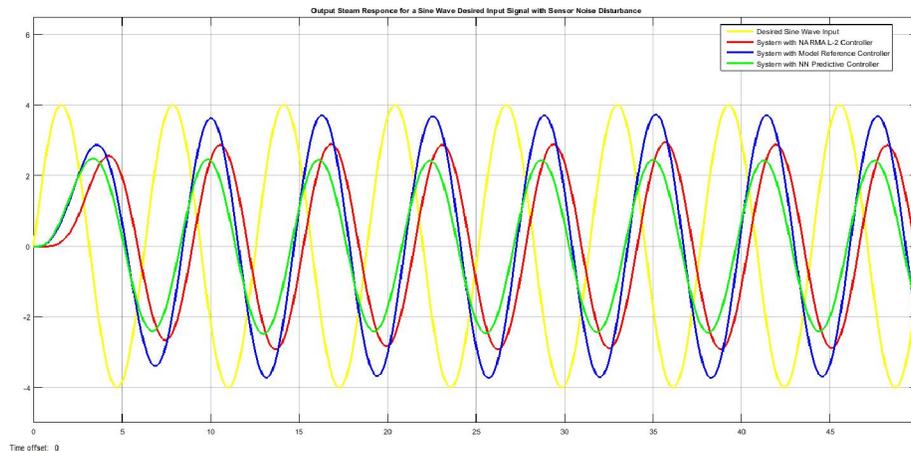

Figure 13 simulation result of the process control system with the proposed controllers for tracking the desired steam input step signal with the presence of steam flow sensor disturbance

Table 5 shows the performance characteristic of the simulation result.

Table 5 Performance characteristic

| No | characteristic | Peak value (m) |
|----|----------------|----------------|
| 1 | Sine wave signal | 4 |
| 2 | NARMA L-2 | 2.8 |
| 3 | Model Reference | 3.7 |
| 4 | NN Predictive | 2.3 |

Table 5 shows that the sensor disturbance affects the tracking progress but still the process control system with model reference controller have track the desired sine wave signal with 3.7 m peak value as compared to the two proposed controllers.

**5.　　　Conclusion**

In this paper, the design and analysis of cement production process control system is done with the help of Matlab/Simulink toolbox successfully. NARMA L2, model reference and neural network predictive controllers are used to improve the system performance for tracking a reference input signal which are step and sine wave. The system is also analyzed when a sensor noise is appearing in the process. From the step response of the system with the proposed controllers, the three controllers have almost the same rise time but the process control system with model reference controller has a small settling time and percentage overshoot as compared to the two proposed controllers and with the presence of sensor noise, the performance characteristic of the three controllers have been changed. But still the process control system with model reference controller has a small settling time and percentage overshoot as compared to the two proposed controllers. From the sine wave response of the system with the proposed controllers, the process control system with model reference controller have track the desired sine wave signal peak value as compared to the two proposed controllers and with the presence of sensor noise, the sensor disturbance affects the tracking progress but still the process control system with model reference controller have track the desired sine wave signal peak value as compared to the two proposed controllers.

Finally the comparison results prove the effectiveness of the presented process control system with model reference controller.

5/24/2020